\documentclass[fleqn,usenatbib]{mnras}

\usepackage{natbib}
\usepackage{newtxtext,newtxmath}

\usepackage[T1]{fontenc}

\DeclareRobustCommand{\VAN}[3]{#2}
\let\VANthebibliography\thebibliography
\def\thebibliography{\DeclareRobustCommand{\VAN}[3]{##3}\VANthebibliography}

\usepackage{graphicx}	
\usepackage{amsmath}	

\usepackage[dvipsnames]{xcolor}
\definecolor{TodoColor}{RGB}{239, 0, 101}

\title[AB Dor's Radio Corona]{Constraining the Coronal Properties of AB Dor in the Radio Regime}

\author[C. E. Brasseur et al.]{
C. E. Brasseur,$^{1}$\thanks{E-mail: cb432@st-andrews.ac.uk}
M. M. Jardine,$^{1}$
G. A. J. Hussain$^{2}$
\\
$^{1}$SUPA, School of Physics and Astronomy, North Haugh, St Andrews, Fife, KY16 9SS, UK\\
$^{2}$European Space Agency, ESTEC, Keplerlaan 1, 2201 AZ Noordwijk, The Netherlands
}

\date{Accepted XXX. Received YYY; in original form ZZZ}

\pubyear{2024}

\begin{document}
\label{firstpage}
\pagerange{\pageref{firstpage}--\pageref{lastpage}}
\maketitle

\begin{abstract}

We present a multiwavelength study of AB Doradus, combining modelling that incorporates a spectropolarimetric magnetic field map with 8.4 GHz radio interferometry to measure the coronal extent and density of this young star. We use the surface magnetic field map to produce a 3D extrapolation of AB Dor's coronal magnetic field. From this model we create synthetic radio images throughout the stellar rotation period which we can compare with the interferometric radio observations. Our models reproduce the two-lobe structure seen in the radio observations. We successfully fit the observed flux magnitude and lobe separation with our model. We conclude that that the features seen in the radio images are a result of centrifugal containment of hot gas at the peak of closed magnetic loops, and that the corona of AB Dor extends to about 8-10 stellar radii, making it much more extended than the present-day solar corona. 
\end{abstract}




\section{Introduction}

The broad strokes of a star's evolution are determined by its mass (as is clear in the standard HR-diagram evolution tracks), but the details of stellar evolution depend also on such properties as magnetic field strength and structure, stellar rotation rate, and chemical composition \citep{Brun_2017LRSP...14....4B}. These properties are themselves also interdependent. In particular the magnetic field of a star allows it to lose angular momentum through the magnetically-channelled stellar wind and thus determines the rate at which it spins down \citep{Weber_1967ApJ...148..217W}. In turn, the rotation rate also governs the magnetic field strength through the action of the dynamo such that when the star rotates faster, the dynamo produces a stronger magnetic field \citep{Hartmann_1985SoPh..100..587H}. How this plays out in the evolution of a particular star also depends on the geometry of the magnetic field and how that affects the stellar wind, mass loss, and related properties. One of the open questions for stars that are young, fast rotators, is the extent of the corona, i.e. how far out in the star's atmosphere are there closed magnetic field lines that can confine the hot coronal gas? We know that the Sun's corona extends to  $\sim2.5 R_\odot$ \citep{Panasenco_2020ApJS..246...54P}, and although we expect a star's magnetic field strength and rotation rate to change the properties of the corona, it is not yet determined how this affects the corona's extent. The Zeeman-Dopper Imaging (ZDI) technique can be used to build surface magnetic field maps, however it is difficult to directly detect magnetic features above the surface of a star. A multi-wavelength approach that combines the surface magnetic map with observations of coronal emission is a very powerful approach.

The character and extent of a star's corona affects not only the star itself, but its immediate environment.  Exoplanets in short-period orbits may pass through both wind and coronal plasmas as they orbit. \citet{Strugarek_2022MNRAS.512.4556S} modelled star-planet interactions for HD 189733 and its planet that orbits at $\sim 9 R_*$;  \citet{Folsom_2020A&A...633A..48F} looked at the closest planet to 55 Cnc, a super-Earth that orbits at just $3.5 R_*$. Both studies predicted that the planets orbit within the coronal extent, crossing both closed magnetic loops and open areas of stellar wind, and both studies discuss the importance of understanding the stellar coronal characteristics when modelling interactions between the orbiting exoplanet and the host star. With an eye to the effects of space weather on orbiting planets, \citet{Davis_Vedantham_2021A&A...650L..20D} studied radio emission from the M dwarf WX UMa and concluded that the closed magnetic field lines extended out to at least $10 R_*$.

One way to quantify the extent of the stellar corona is define a theoretical radius beyond which all field lines become radial. This radius is called the  ``source-surface'' ($R_{ss}$) location,  and while this imaginary ``surface'' is not in reality spherical, it is a useful construct. \citet{Reville_2015ApJ...814...99R} developed a criterion for estimating the ``source-surface'' ($R_{ss}$) location. \citet{See_2018MNRAS.474..536S} used this method to extrapolate the optimal source-surface location for 22 solar analogue stars with ZDI magnetic field maps, developing source-surface evolution tracks for three stellar rotation rate.

In this paper we will be studying AB Doradus A (hereafter AB Dor), one of the best observed stars in the \citet{See_2018MNRAS.474..536S} sample. AB Dor is the main star in a quadruple system consisting of two gravitationally associated binaries \citep{Guirado_2006A&A...446..733G}.  It is a fast-rotating (0.514 day) zero-age main sequence star of stellar type K1. AB Dor has the advantage of being well studied over decades with a variety of techniques and as many wavelengths, making it an excellent candidate for detailed study as there are good constraints on its directly observable quantities. A selection of AB Dor's properties relevant to this study are presented in Table \ref{tbl:ab_dor}. AB Doradus Ba/Bb are a close binary companion to AB Doradus A/C with a 9 arcsec separation (135 AU) \citep{Guirado_2006A&A...446..733G}. This is distant enough that while the two binaries are gravitationally linked they do not directly affect one another and can be analysed separately. AB Doradus C is a very low mass companion to AB Doradus A. Because of its closeness (4.5 AU), AB Doradus C cannot be dismissed in analyses of AB Dor A observations, however it is too small ($\sim 0.09 M_\odot$) to affect the gravitational potential of AB Dor A, so we do not include it in our model \citep{Guirado_2006A&A...446..733G}.

{\small 
\begin{table}
    \centering
    \begin{tabular}{ccc}
        \textbf{Property} & \textbf{Value} & \textbf{Source} \\
       \hline
       Mass & $0.89 \pm 0.08 M_\odot$ & [1]\\ 
       Distance & 14.9 pc & [2]\\
       Rotation Period & 0.514 day & [3]\\
       Age & 40-50 Myr & [1]\\
       Radius & $0.96 \pm 0.06 R_\odot$ & [4]\\
       Corotation radius ($r_K$) & $2.6R_*$ & [5]\\
       \hline
        \multicolumn{3}{l}{[1] \citet{Azulay_2017A&A...607A..10A}} \\
        \multicolumn{3}{l}{[2] \citet{Guirado_2006A&A...446..733G}} \\
        \multicolumn{3}{l}{[3] \citet{Innis_1985PASA....6..160I}} \\
        \multicolumn{3}{l}{[4] \citet{Guirado_2011A&A...533A.106G}} \\
        \multicolumn{3}{l}{[5] \citet{Villarreal_2018MNRAS.475L..25V}} \\
    \end{tabular}
    \caption{Stellar properties of AB Doradus A, referred to throughout this paper as AB Dor. }
    \label{tbl:ab_dor}
\end{table}
}

AB Dor is a well-known flare star; X-ray flare monitoring indicates that it flares nearly once a day \citep{Hussain_2007MNRAS.377.1488H}, and optical observations show  a ``superflare'' event ($>10^{34}$ erg) about once a week \citep{Schmitt_2019A&A...628A..79S}. This indicates a very active corona, significantly more so then our own Sun.  \citet{Lalitha_2013A&A...559A.119L} investigated the possibility of an X-ray activity cycle counterpart to the observed $\sim 17$ year photospheric activity cycle in AB Dor. They found that the X-ray brightness varies within in a much smaller range than  that of the optical V-band, showing limited long term variation in contrast to the high rate of short term variation caused by continuous flaring activity \citep{Maggio_2000A&A...356..627M}. X-ray studies also indicate a that AB Dor has a hot corona, on the order of 8-10 MK (see \citet{Hussain_2005ApJ...621..999H, Close_2007ApJ...665..736C}).

Stellar ``slingshot prominences'' were first observed on AB Dor in the H$\alpha$ band. \citet{Cameron_1989MNRAS.236...57C} identified a stellar analogue to solar prominences (cooler, denser clouds of gas, trapped in pressure maxima within the corona), noting that on AB Dor these features were more massive and higher in the corona than on the Sun. Many studies of stellar prominences followed this one (e.g. \citet{Jardine_1991SoPh..131..269J, Jardine_2019MNRAS.482.2853J, Villarreal_2018MNRAS.475L..25V}). Prominences are seen in absorption (usually Balmer lines) as they traverse in front of the star, or in emission when they are beyond the stellar limb \citep{ Dunstone_II_2006MNRAS.373.1308D,Villarreal_2019MNRAS.485.1448V}. Their radial accelerations have been observed to indicate locations near the Keplerian co-rotation radius ($r_K\sim2.6R_*$ for AB Dor).  While the discovery of prominences orbiting in the stellar corona near the co-rotation radius have made it clear that AB Dor's corona may be larger than the Sun's, its extent has not been fully determined. Given AB Dor's age (40-50 Myr), \citet{See_2018MNRAS.474..536S} predict  a very large source-surface, anywhere from $11-27 R_*$ depending on the assumed rotational evolution track.

AB Dor's surface magnetic field was first mapped in 1995 \citep{Donati_1997MNRAS.291....1D},  and between then and 2007 it was mapped nearly annually.   \citet{Hussain_2007MNRAS.377.1488H} used contemporaneous X-ray light curves and ZDI magnetic surface maps to correlate surface activity with the X-ray corona. In contrast to the H$\alpha$ prominence studies, they determined that the X-ray corona of AB Dor must be very compact ($\sim 0.3-0.4 R_*$). In additional evidence of a compact X-ray corona, \citet{Maggio_2000A&A...356..627M} studied two large X-ray flares, and concluded that they could not originate in AB Dor's prominences, but instead must come from coronal loops with maximum heights on the order of $1 R_*$.  \citet{Jardine_2005MNRAS.361.1173J} constructed a model for AB Dor that reconciled the compact nature of the observed X-ray corona with the extended features observed in H$\alpha$.  In their model, gas that was originally part of the stellar wind is trapped in larger closed magnetic loops that are not X-ray bright, but can support the prominences. When the prominence mass can no longer be supported by the magnetic field, the prominence either falls back to the surface (if it lies below $r_K$) or is centrifugally ejected (if it lies above $r_K$).  These latter prominences are the so-called ``slingshot prominences'' that carry away mass and angular momentum from the star \citep{Villarreal_2018MNRAS.475L..25V}. 

Evidence of a compact X-ray corona existing within a more extended radio corona is discussed in the \cite{Massi_2008A&A...480..489M} observations of binary system V773 Tauri. In that study they specifically consider and find evidence for solar-like ``helmet streamers.'' These are magnetic structures that form at the top of X-ray emitting coronal loops such that there is an extended magnetic ``streamer'' extending many stellar radii and facilitating the flow of plasma between upper and lower ``mirror point'' at the peak of the helmet streamer, and anchor point where it attaches to the coronal loop \citep{Massi_2006A&A...453..959M}. While these structures have not been observed on AB Dor (and indeed further observations could not confirm their presence on V773 Tauri \citep{Torres_2012ApJ...747...18T}), the evidence for a compact X-ray corona, extended radio corona, and the presence of prominences makes AB Dor a candidate to host them \citep{Climent_2020A&A...641A..90C}. Large-scale prominence-like structures were also detected in time-series of HST GHRS far-ultraviolet spectra of the eclipsing binary system, V471 Tau \citep{Walter_2004AN....325..241W}.
 
The radio emission from AB Dor was studied in 1994 by \citeauthor{Lim_1994ApJ...430..332L}. Outside of impulsive flaring events they observed no circular polarisation and minimal variability. They based their observations on radio light curves, and after phase-folding were able to see radio peaks consistent with star spots. More recently radio observations of the AB Doradus system have been used to constrain the dynamical masses of all four members (e.g. \citet{Wolter_2014A&A...570A..95W,Azulay_2015A&A...578A..16A,Azulay_2017A&A...607A..10A}). Even more recently, in 2020, \citeauthor{Climent_2020A&A...641A..90C} produced a new study of AB Dor in the radio regime. They used very long baseline interferometry (VLBI) to probe the magnetosphere of AB Dor.  The images they produced show complex structures around the star, with two distinct emission lobes. These lobes vary in position and strength on a timescale of hours, making them likely related to the rotation of the star. \citet{Climent_2020A&A...641A..90C} put forth four possible scenarios compatible with their observations. Motivated by this work we use a ZDI surface magnetic field map to build a suite of 3D models of the coronal magnetic field, and explore how the properties of our model are reflected in synthetic radio images created from it. \citet{Climent_2020A&A...641A..90C} present VLBI data from a number of years between 2007 and 2018; we focus on their 2007 data because there is also a ZDI map from that year.

\section{Methods}\label{sec:methods}

\subsection{Coronal Magnetic Field Model}\label{sec:magfield_model}

The Zeeman-Doppler Imaging (ZDI) technique is a tomographic technique that uses time series spectropolarimetric observations over a full rotation period to recover the large-scale magnetic field strength and geometry of a star \citep{Semel_1989A&A...225..456S}.  We use the radial field component of the Zeeman-Doppler maps described in \citet{Cohen_2010ApJ...721...80C}  to extrapolate the magnetic field structure of the corona between the surface of the star and the ``source surface,'' the imaginary shell at the location where all magnetic field lines become radial. 

The method we use to extrapolate the coronal field is described in detail in \citet{Jardine_2002MNRAS.333..339J}, implemented in code originally developed by \citet{vanBallegooijen_1998ApJ...501..866V}.  Broadly speaking, we assume that the stellar magnetic field is ``potential,'' (the curl of the magnetic field is zero), so we can describe the magnetic field ($\mathbf{B}$) in terms of a scaler potential ($\Psi$), such that $-\nabla \Psi = \mathbf{B}$. The requirement that the magnetic field is divergence free then reduces the problem to Laplace's equation $\nabla^2\Psi=0$, which we solve in spherical coordinates.  The potential field assumption is appropriate for our model because we are interested in the large-scale structure, which is well reproduced by this approach \citep{Riley_2006ApJ...653.1510R}. Additionally,  when \citet{Jardine_2013MNRAS.431..528J} explored the effect of the non-potential component of the stellar surface magnetic field they concluded that the non-potential nature of the field does not significantly affect the coronal wind.

Once we have the magnetic field model for the corona (i.e. the magnetic field $\mathbf{B}$ at every point in the spherical grid), we trace individual field lines through each grid cell, collecting an array of open and closed field lines, each discretised into an array of points with a given position ($\theta, \phi, r$), volume, cross-sectional area, magnetic field strength, pressure, and number density. When tracing the field lines we make the assumption of an isothermal corona at temperature $T$ in hydrostatic equilibrium.  With these assumptions in place we can write the pressure at any point along a given field line as
\begin{equation}
    p = p_0e^{(m/kT)\int{g_s}ds}
\end{equation}
where $g_s$ is the component of the effective gravity along the field line ($g_s = \mathbf{g}\cdot \mathbf{B}/|\mathbf{B}|$),  $p_0$ is the pressure at the field line footpoint, $m$ is the mean particle mass, and $k$ is the Boltzmann constant. We relate the plasma pressure at the footpoint to the magnetic pressure with
\begin{equation}
    p_0(\theta,\phi) \propto B_0^2 (\theta,\phi),
\end{equation}
which is a  formulation presented in \citet{Jardine_2002MNRAS.333..339J}. We can then tune the relation with a multiplier on $B_0^2$. This allows us to set the base pressure of the corona as part of our model.

Once we have a list of  field lines, we use the method outlined in \citet{Jardine_2020MNRAS.491.4076J} to  search all closed field lines for stable points where prominences are able to form. We do this by searching for pressure maxima, i.e. points on the field line that satisfy the condition
\begin{equation}
(\mathbf{B} \cdot \mathbf{\nabla})(\mathbf{g} \cdot \mathbf{B}) = 0
\end{equation}
where $\mathbf{B}$ is the magnetic field and $\mathbf{g}$ is the effective gravity (see \citet{Ferreira_2000MNRAS.316..647F} for a detailed discussion of the stability criterion). We then fill the field line flux tube with the maximum amount of mass that can be supported at the prominence location, given a specified prominence temperature ($8500$ K for AB Dor \citep{CollierCameron_2001IAUS..203..229C}). Once the prominence mass has been determined it is allowed to spread out hydrostatically along the field line. The affected field line cells are then marked as being part of a prominence, and the density and pressure updated accordingly.

\begin{figure}
 \centering
    \includegraphics[trim={2.5cm .5cm 2.5cm .5cm},clip, width=\columnwidth]{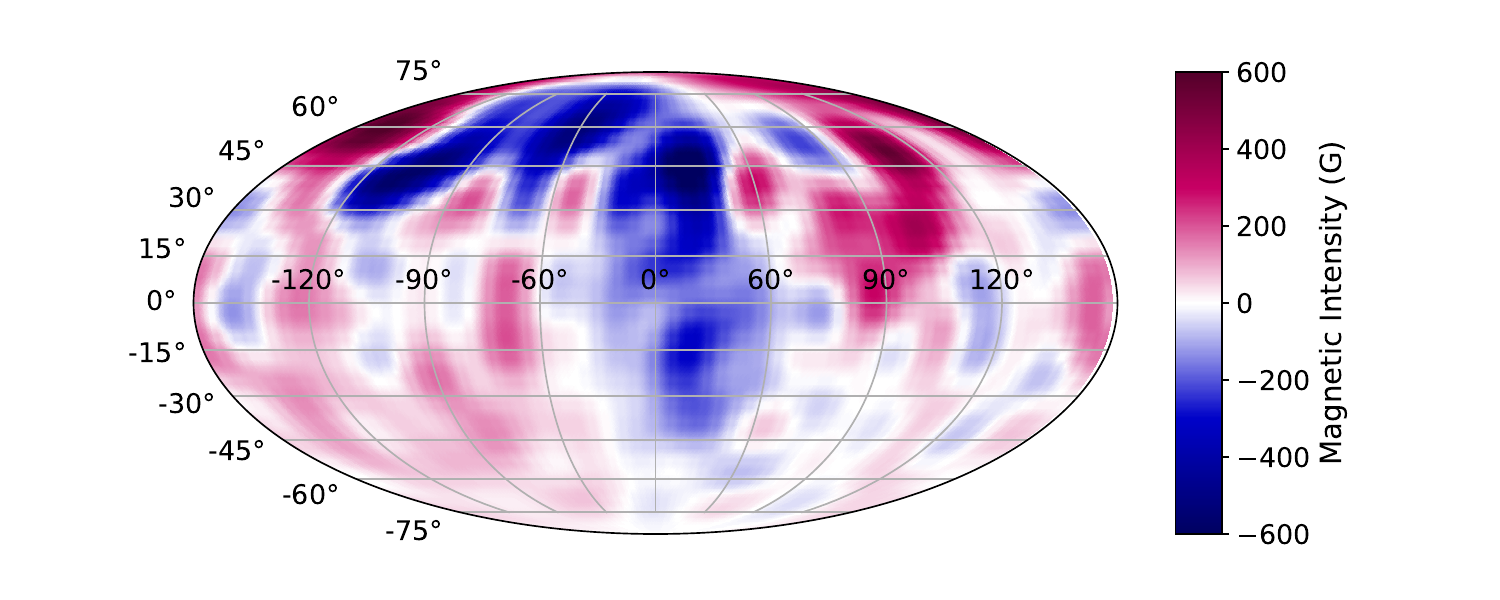}
    \caption{The radial component of the ZDI-determined surface magnetic field map for AB Dor from December 2007 \citep{Cohen_2010ApJ...721...80C}.
    \label{fig:zdi_map} }
\end{figure}

The ZDI surface magnetic field map we use is that from \cite{Cohen_2010ApJ...721...80C}, which was built using the technique described in \citet{Hussain_2002ApJ...575.1078H}. Because AB Dor's rotation axis is inclined with respect to the line of sight to the observer, some of the AB Dor surface cannot be observed. The particular map we use is the ``unconstrained'' ZDI solution, where the maximum entropy solution is preferred with no additional guiding principles for the areas of the surface that we cannot observe (e.g. that the magnetic field should be symmetric or antisymetric, see \citet{Donati_1997MNRAS.291....1D, Donati_1999MNRAS.302..437D,Hussain_2007MNRAS.377.1488H, Lehmann_2019MNRAS.483.5246L}). The surface magnetic field map that forms the basis of our models is shown in Figure \ref{fig:zdi_map}. The three dimensional grid we build when we extrapolate the coronal field is linear in $\theta$ and $\phi$, with a resolution of $64 \times 128$, and exponential in radius such that for a radial shell $i$ (where $i = 0$ is the stellar surface), the normalised radius of points on that shell ($r_i = R_i/R_*$) are related to the surface resolution by $r_i = e^{i\pi/N_\theta}$, where $N_\theta = 64$ is the resolution in $\theta$.

\begin{figure}
 \centering
    \includegraphics[trim={1.5cm 1cm 1.5cm 2cm},clip, width=\columnwidth]{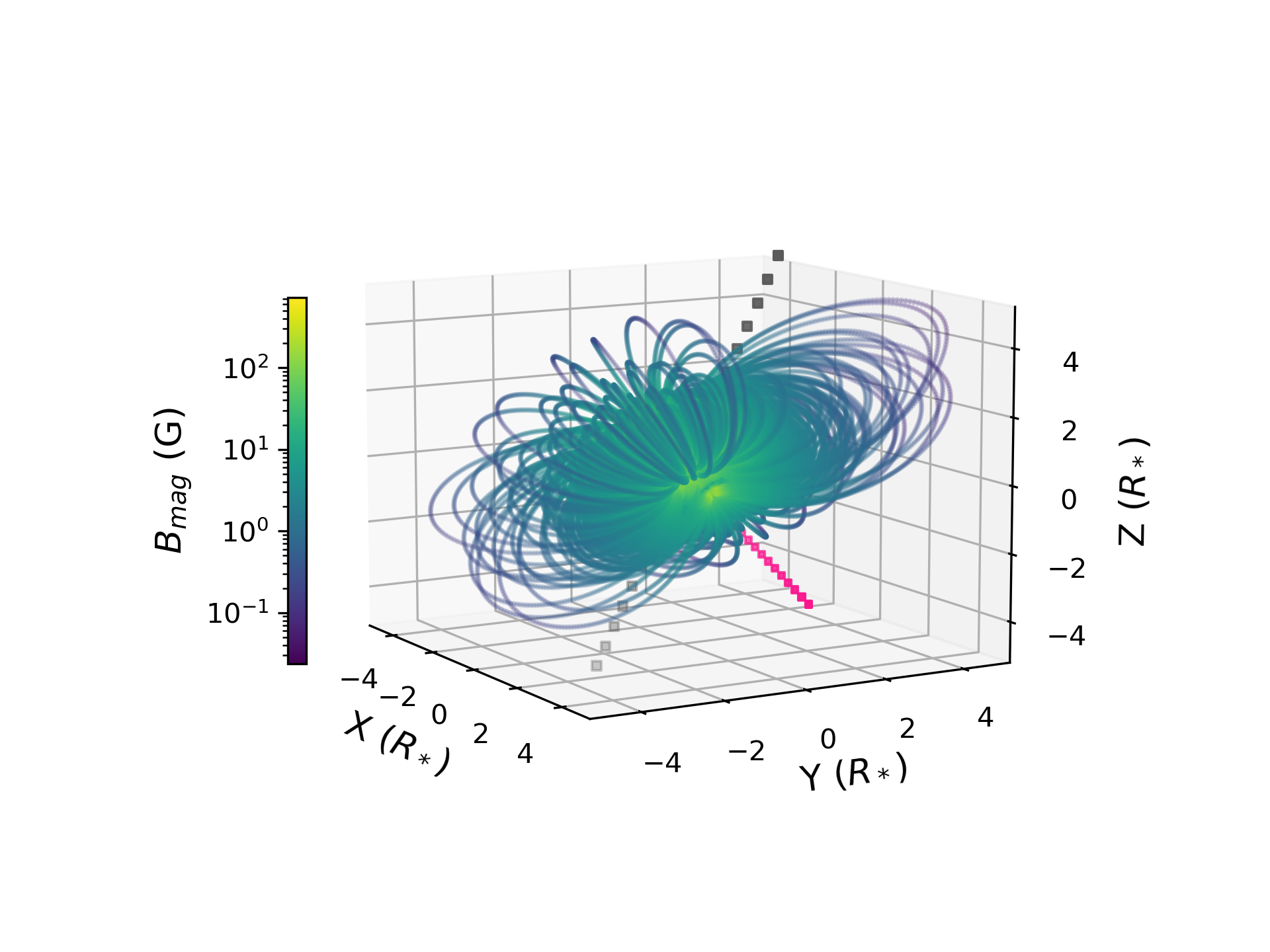}
    \includegraphics[trim={1.5cm 1cm 1.5cm 2cm},clip, width=\columnwidth]{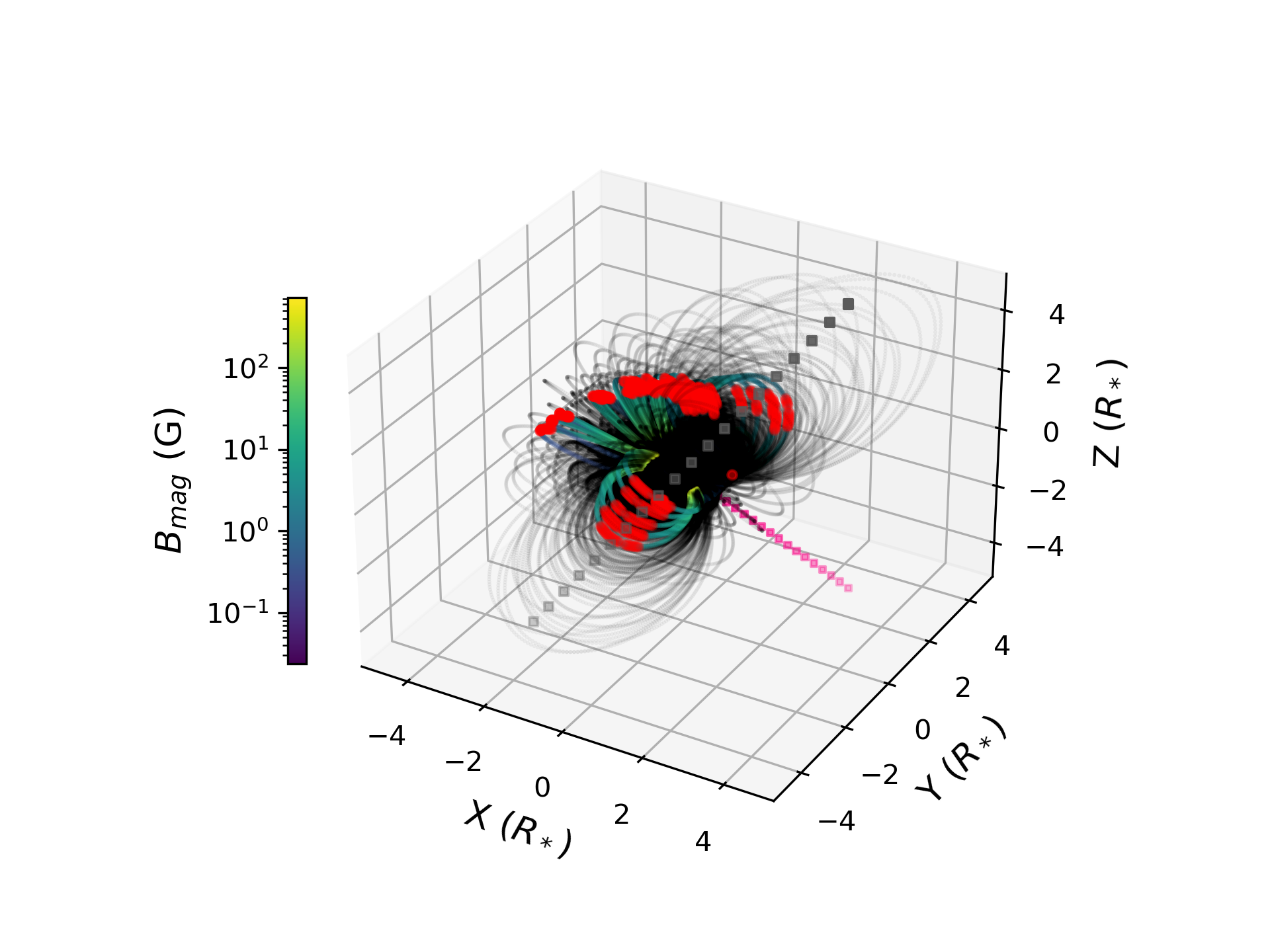}
    \caption{AB Doradus model magnetic field lines, given the parameters in Table \ref{tbl:model_parms}. The \textit{top} plot shows all the closed field lines coloured by magnetic field strength. The \textit{bottom} plot focuses on the prominence bearing field lines, which are still coloured by magnetic field strength while the non-prominence bearing lines are picked out in black, and the prominences themselves are shown in red. The dotted grey line in both plots is the axis of rotation, and the dotted pink line is the magnetic dipole axis. The rotation axis is tilted $60
^\circ$ from our viewing angle, and the dipole axis  is misaligned from the magnetic field axis by about $100^\circ$.
    \label{fig:field_lines} }
\end{figure}

Figure \ref{fig:field_lines} shows our magnetic field extrapolation given the parameters listed in Table \ref{tbl:model_parms}. These parameters are used for the models shown in figures throughout the paper unless otherwise noted. The top image shows all of the closed field lines coloured by magnetic field strength.  The bottom image shows only the prominence-bearing field lines coloured by magnetic field strength, with the prominences themselves marked in red, and the non-prominence-bearing field lines picked out in black. The rotation axis is marked by the grey  dotted line, while the magnetic dipole axis is in pink. We can see that while there are closed field lines extending out to the source surface, the prominences cluster nearer the star, around the co-rotation radius.

{ 
\begin{table}
    \centering
    \begin{tabular}{rl}
    \hline
        \textbf{Property} & \textbf{Value}  \\
       \hline
        ZDI map year$^{\dagger}$  & 2007\\
        Prominence temperature ($T_{prom}$)$^{\ddagger}$ & $8,500~\mathrm{K}$\\
        \hline
        Base pressure  ($p_0$) & $10^{-5.5}\times B_0(\mathrm{G})^2$ Pa\\
        Source surface radius ($R_{ss}$) & $8.1 R_\odot$\\
        Corona temperature ($T_{cor}$) & $2 \times 10^6~\mathrm{K}$\\
       \hline
       Observation frequency & 8.4 GHz\\
       Resolution & .01$R_*/$ px \\
       Image size & $200 x 200$ px \\
       \hline
       \multicolumn{2}{l}{$^{\dagger}$ \citet{Cohen_2010ApJ...721...80C} } \\
 \multicolumn{2}{l}{$^{\ddagger}$ \citet{CollierCameron_2001IAUS..203..229C}} 
    \end{tabular}
    \caption{Model values used in producing the figures throughout this paper unless otherwise noted.}
    \label{tbl:model_parms}
\end{table}
}

\subsection{Synthetic Image creation}\label{sec:image_creation}

\begin{figure}
 \centering
    \includegraphics[width=\columnwidth]{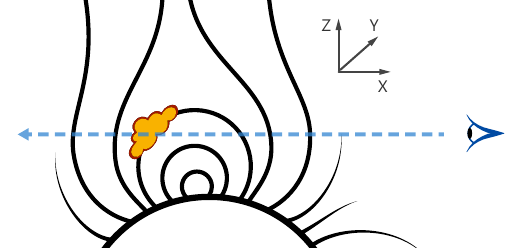}
    \caption{This diagram illustrates an example sight-line through the stellar corona. The sight-line crosses both open and closed field lines, as well as a prominence.
    \label{fig:prominence_diagram} }
\end{figure}

Once we have our 3D model of the coronal magnetic field, we use it to produce synthetic images at a particular observation wavelength, viewing angle, and rotation phase. Figure \ref{fig:prominence_diagram} shows an example sight line through a stellar corona including both closed and open field lines, and a stellar prominence. 

We choose an observational wavelength of 8.4 GHz, the frequency at which \citet{Climent_2020A&A...641A..90C} observed during the year we have a ZDI map. In the hot, fully-ionised corona of our star AB Dor, we expect that the quiet background emission will come from free-free (bremsstrahlung) emission as it does on the Sun \citep{Dulk_2001pre5.conf..429D}. This  emission is incoherent and often optically thick, so it is important to include the optical depth when modelling this emission \citep{Dulk_1985ARA&A..23..169D}. The only line emission present at 8.4 GHz are radio recombination lines, however AB Dor's corona is too hot for this line emission which is visible at temperatures $\sim 10^4$ K \citep{Dravskikh_2022ARep...66..490D}.

Because this is a static model, transient effects such as flares, are also not included.  While AB Dor is a well known flare star,  \citet{Climent_2020A&A...641A..90C} show that the morphology of the radio emission is persistent over the 10 years of their observations, and thus we find it reasonable to search for physical underpinnings to the radio morphology in the steady-state realm. We also do not model emission from gyrosynchrotron, cyclotron maser, and plasma radiation, as those emission mechanisms are all associated with short term activity \citep{Dulk_2001pre5.conf..429D}. That the observed emission is only weakly polarised, supports this decision.

We do not include emission from the stellar wind (open magnetic field lines) in our calculations because its contribution is negligible.  Using the given stellar wind density and temperature approximations from \citet{OFionnagain_2018MNRAS.476.2465O}, we estimated the wind emission using a simple 1-dimensional Parker Wind solution and found that the total emitted flux ranged from a factor of $\sim 10^{-4}-10^{-1}$ of the flux from the closed corona.

Our method of calculating the optical depth is based on that of \citet{Wright_1975MNRAS.170...41W}, who presented a model for the radio and infrared spectrum of early-type stars, based on the assumption of uniform mass loss. The original formulation of this model required the stellar wind to be assumed spherically symmetrical and  at terminal velocity, \citet{DaleyYates_2016MNRAS.463.2735D} used numerical methods to relax these constraints, and used the resulting algorithm to create synthetic radio emission observations for accelerating stellar winds. We follow the \citet{DaleyYates_2016MNRAS.463.2735D} method, but further relax the constant temperature constraint to account for the presence of stellar prominences.

As discussed in the previous section we model the corona as isothermal ($T_{c}$) and in hydrostatic equilibrium, with prominences at a single cooler temperature ($T_{p}$). We model both the corona and prominences as fully ionised hydrogen, emitting Blackbody continuum radiation at their given temperatures. This assumption of full ionisation is very close to the true conditions in the corona, which is hot enough to be fully ionised and is nearly entirely made up of hydrogen. For the prominences, however, while they have the same atomic makeup as the corona, because they are cooler and denser a smaller proportion of the gas will be ionised. The neutral gas in the prominences does not however affect the ability of the magnetic field to confine the prominence gas. This was determined in studies of the Sun such as \citet{Pneuman_1971SoPh...18..258P}. 

Given our assumptions, the intensity of the emission at every point in the (closed magnetic field) corona can be expressed by  Planck's law (which we use in its full form rather than the Rayleigh–Jeans approximation). Planck's law describes the radiation as it is emitted from each individual location, however we must also consider the opacity of the coronal gas and prominences, meaning that the intensity that reaches the observer will be
\begin{equation}
    I_\nu(y,z) = \int_{-\infty}^{\infty}B_\nu \left ( T \left ( x,y,z \right ) \right )~e^{-\tau(x,y,z)}\kappa_{ff}(x,y,z)~\mathrm{d}x
\end{equation}
where $I_\nu(y,z)$ is the intensity at position $(y,z)$ in the observation plane, $B_\nu$ is the Blackbody emission at position $(x,y,z)$ in the corona for a observing frequency $\nu$,  $\tau$ is the optical depth at position $(y,z)$  and distance $x$ along the line of sight, and $\kappa_{ff}$ is the free-free absorption coefficient \citep{DaleyYates_2016MNRAS.463.2735D}, see Figure \ref{fig:prominence_diagram}. The integral is defined over the entire line of sight; in practice this means we integrate across the volume of our model.

As we are only modelling bremsstrahlung emission, the optical depth can be related to the free-free absorption coefficient with
\begin{equation}
    \mathrm{d}\tau = \kappa_{ff}\mathrm{d}x
\end{equation}
where $\mathrm{d}\tau$ is the infinitesimal optical depth of the material across $\mathrm{d}x$. 

While the stellar prominences are not completely ionised, at radio frequencies the neutral atoms do not contribute to absorption, and at the typical densities of stellar prominences the resulting absorption is negligibly affected by even very low ionization fractions. \citet{Mihalas_1978stat.book.....M} derives the free-free absorption coefficient using the quantum defect method, assuming local thermodynamic equilibrium and a fully ionised 100\% hydrogen atmosphere. The result in CGS units is
\begin{equation}
    \kappa_{ff} = 0.0178\frac{Z^2g_{ff}}{T^{3/2}\nu^2}n_e n_i \label{eqn:kff}
\end{equation}
where $Z$ is the ion charge, $g_{ff}$ is the free-free Gaunt factor, $T$ is the temperature, $\nu$ is the observing frequency, and $n_e$ and $n_i$ are the electron and ion number densities respectively. We use the free-free Gaunt factor from \citet{DaleyYates_2016MNRAS.463.2735D}, derived by \citet{vanHoof_2014MNRAS.444..420V} 
\begin{equation}
    g_{ff} = 9.77 + 1.27\log_{10}\left( \frac{T^{3/2}}{\nu Z} \right )
\end{equation}
where all of the symbols are the same as in equation \ref{eqn:kff}.

Combining these three equations, for a given viewing angle (see Figure \ref{fig:prominence_diagram}), we calculate the optical depth along a particular sight-line (in the $\hat{x}$ direction) based on the free-free absorption coefficient ($\kappa_{ff}$) and the opacity of the material between the viewer and the $x-$axis position. Thus
\begin{equation}
    I_\nu = \sum^j B_\nu(T_j) e^{-\tau_j} \kappa_{ff,j} \mathrm{d}x_j 
\end{equation}
and,
\begin{equation}
    \tau_j = \sum_{i=0}^{i=j} \kappa_{ff,i}\mathrm{d}x_i 
\end{equation}
where $I_\nu(y,z)$ is the intensity at frequency $\nu$ for the sight-line defined by the Cartesian position $(y,z)$, $B_\nu(T_j)$ is the Blackbody radiation in frequency $\nu$ for the temperature $T$ at position $x_j$, and $\tau_j$ is the optical depth at position $x_j$.

To create a synthetic image we first choose the observation angle, as well as the desired image resolution. By default the model is constructed such that the rotation axis points in the $\hat{z}$ direction with rotational phase $\phi = 0^\circ$, however we can rotate the model in any direction by performing a change of coordinate function to simulate any viewing angle, and then rotate the star with $\phi$.

We rotate our model to the desired viewing angle, and then interpolate onto a Cartesian grid at the specified resolution. The interpolation is done using the nearest neighbour method. As mentioned previously, we do not consider the stellar wind in this model, and thus set the density along the open field lines to zero before interpolation. Once we have a 3D Cartesian grid aligned with the line-of-sight, we can numerically integrate through the cube to get first the optical depth in each grid cell, and then the intensity. The last step necessary for comparing our synthetic images with real data is to translate the calculated intensity into observed flux. The spectral flux density ($S_\nu$) at distance $D$ is given by
\begin{equation}
    S_\nu = \int_{-\infty}^{\infty}\int_{-\infty}^{\infty} I_\nu(y,z) \mathrm{d}y\mathrm{d}z.
\end{equation}
For each individual pixel this becomes
\begin{equation}
    S_\nu(y,z) = I_\nu(y,z) \mathrm{d}y\mathrm{d}z
\end{equation}
where $\mathrm{d}y,~\mathrm{d}z$ are the dimensions of the pixel, and the total flux is
\begin{equation}
    S_\nu = \sum^j\sum^i I_\nu(y_j,z_i) \mathrm{d}y_j\mathrm{d}z_i.
\end{equation}

We know that the rotation axis of AB Dor is inclined by $60^\circ$ to our line of sight, so we choose a viewing angle of $(60^\circ, 0^\circ$), where $60^\circ$ is the angle of inclination, and $0^\circ$ is the position angle of the rotation axis in the plane of the sky. While the position angle of AB Dor is unknown, changing it simply causes a rotation of the resulting image, which does not affect our conclusions.

\begin{figure}
 \centering
    \includegraphics[width=0.7\columnwidth]{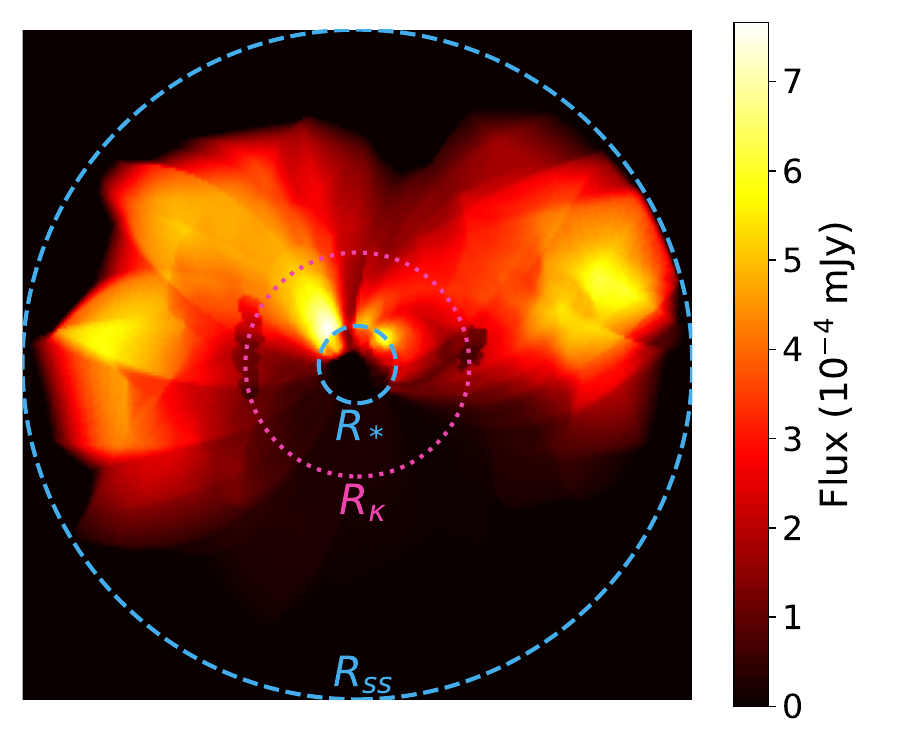}
    \caption{Synthetic image of AB Dor at 8.4 GHz, made using the parameters shown in Table \ref{tbl:model_parms}. The inner blue dashed circle marks the stellar radius, and the outer circle the source surface radius, which is also the edge of the model. The pink dashed circle marks the co-rotation radius. Note the dark features at the co-rotation radius: these are the stellar prominences that cluster around that radius and are radio dark and optically thick.
    \label{fig:flux_img} }
\end{figure}

Figure \ref{fig:flux_img} shows an $8.4$ GHz image produced in this manner with a $60^{\circ}$ axial tilt, at rotation phase $0^{\circ}$, and with an image resolution of $0.1 R_*$ or 200 pixels per side.

\begin{figure}
 \centering
    \includegraphics[width=\columnwidth]{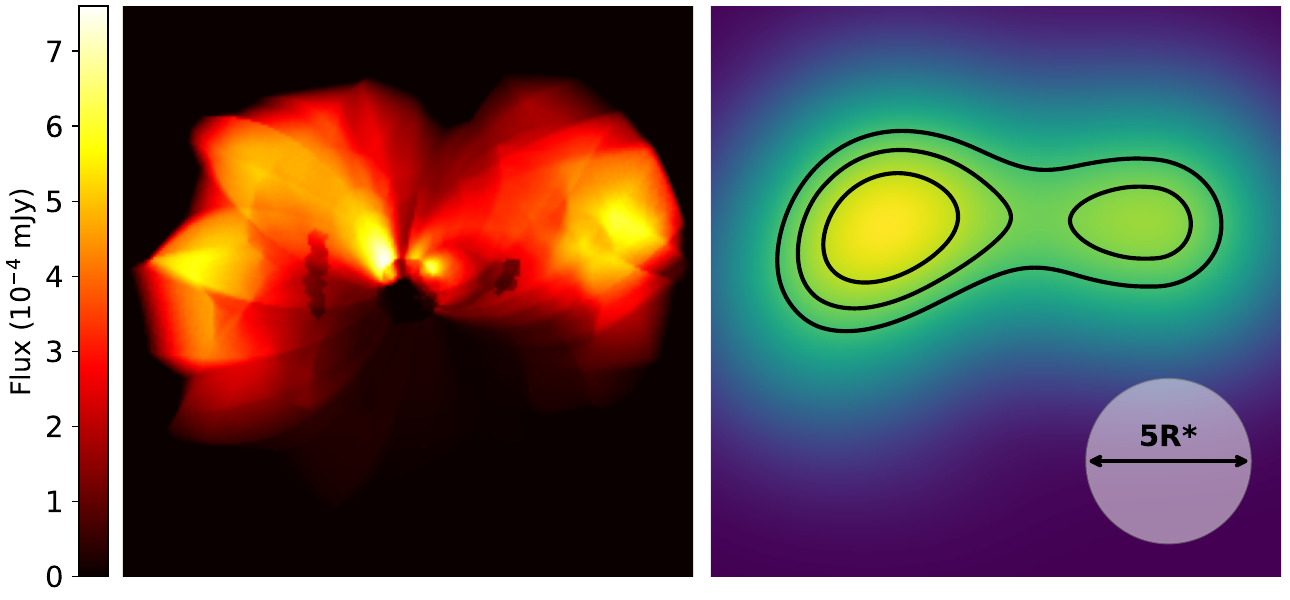}
    \caption{Comparison of the high resolution synthetic image (\textit{left}), and synthetic image convolved with the telescope beam size (\textit{right}) for the parameters given in table \ref{tbl:model_parms}. The circle in the right plot shows the beam size.
    \label{fig:lobe_ex} }
\end{figure}

In the resulting synthetic image (Figure \ref{fig:flux_img}) we can see qualitatively the two lobe structure discussed by \citet{Climent_2020A&A...641A..90C}, however the resolution is clearly much higher than is detectable by the telescope. We therefore convolve the image with the approximate beam size of the instrument, and apply a peak-finding algorithm.  The beam size for the 2007 images in \citet{Climent_2020A&A...641A..90C} is $1.7\times2.7$ mas. Because we do not know the position angle on the plane of the sky for the image, we convolve the image with a circular beam of the smaller diameter. Figure \ref{fig:lobe_ex} shows the high resolution image beside the convolved image with $70\%$, $80\%$, and $90\%$ contours marked. On the convolved image we can see the two lobe structure much more clearly, and measure the distance between the two visible emission peaks. We calculate not only the geometric distance between the emission peaks, but also the distance from the star of each peak and the angular distance between them. If there is only one peak detected we consider there to be no distinct lobes and record a lobe separation of 0. In cases where there are more than one peak we record all of them, but consider the greatest distance between any pair (geometric or angular) to be the distance for the image. We do this to ensure that we record the outermost visible lobes in the corona, giving us the full extent over which we observe flux.

\section{Results}

\begin{figure}
 \centering
    \includegraphics[width=\columnwidth]{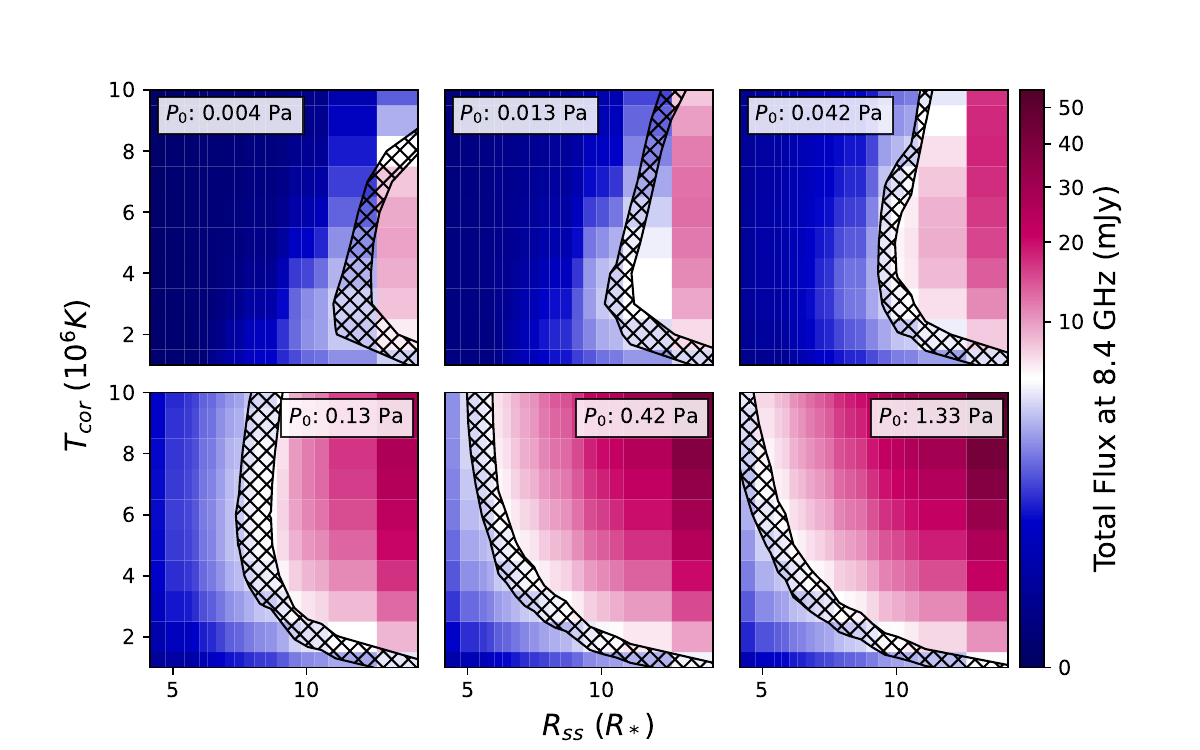}
    \includegraphics[width=\columnwidth]{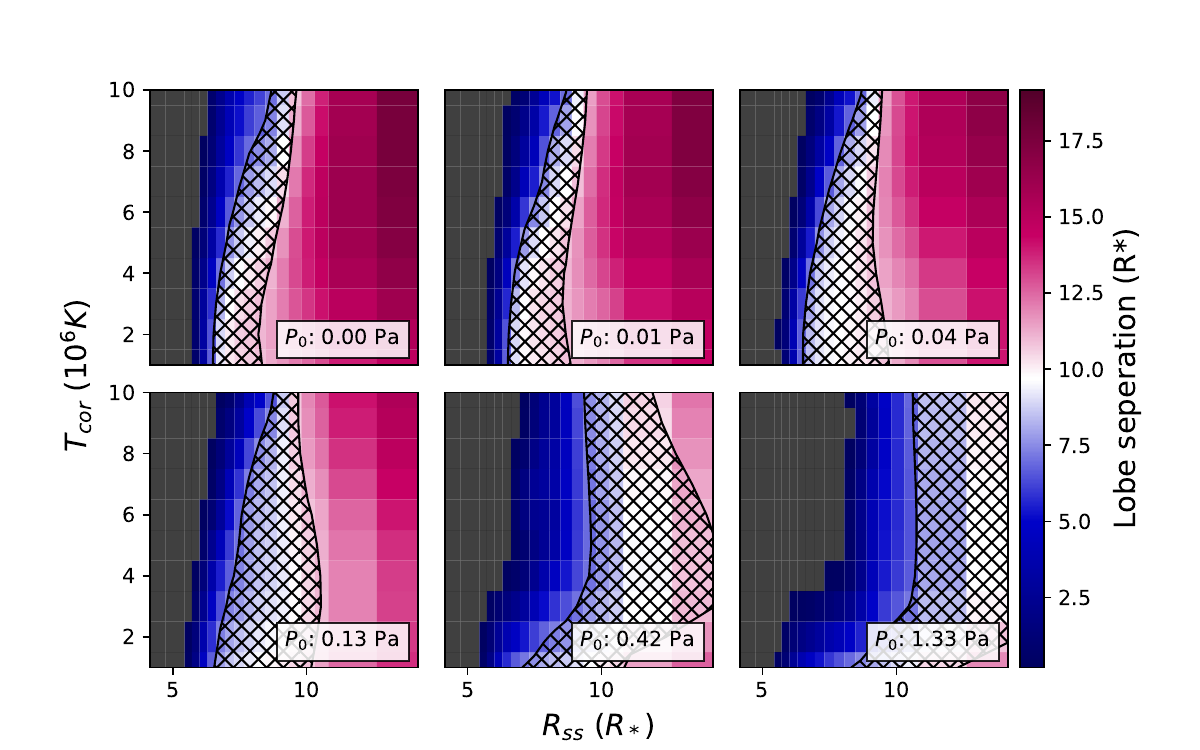}
    \caption{Two plots illustrating our model parameter space. Both plots are coronal temperature vs. source surface radius for six different mean surface pressures. The top plot is coloured by total flux (averaged over the stellar rotation), and the bottom by lobe separation (averaged over the fraction of the rotation period showing two-lobe structure). The black areas in the lower plot indicated models that never show a two-lobe structure. The hatched areas guide the eye to the model space most consistent with the observations in \citet{Climent_2020A&A...641A..90C}; flux values between $3.7-5.7$ mJy, and separations of $7-11 R_*$.  \label{fig:t_vs_rss}}
\end{figure}

Figure \ref{fig:t_vs_rss} shows the outputs of our model as we vary the input parameters. The three free parameters in our model are source surface radius (x-axis), coronal temperature (y-axis), and base pressure (constant in each subplot). The top plot is coloured by the total flux in our synthetic images  (averaged over rotation phase), and the bottom figure is coloured by separation of the two emission peaks. For the emission peak separation we average over all rotation phases that show two peaks of emission in the convolved image; models where no rotation phase results in two emission peaks are coloured black.  The hatched contours guide the eye to the parts of parameter space where the model outputs  most closely align with the  \citet{Climent_2020A&A...641A..90C} results.

\begin{figure}
 \centering
    \includegraphics[width=.9\columnwidth]{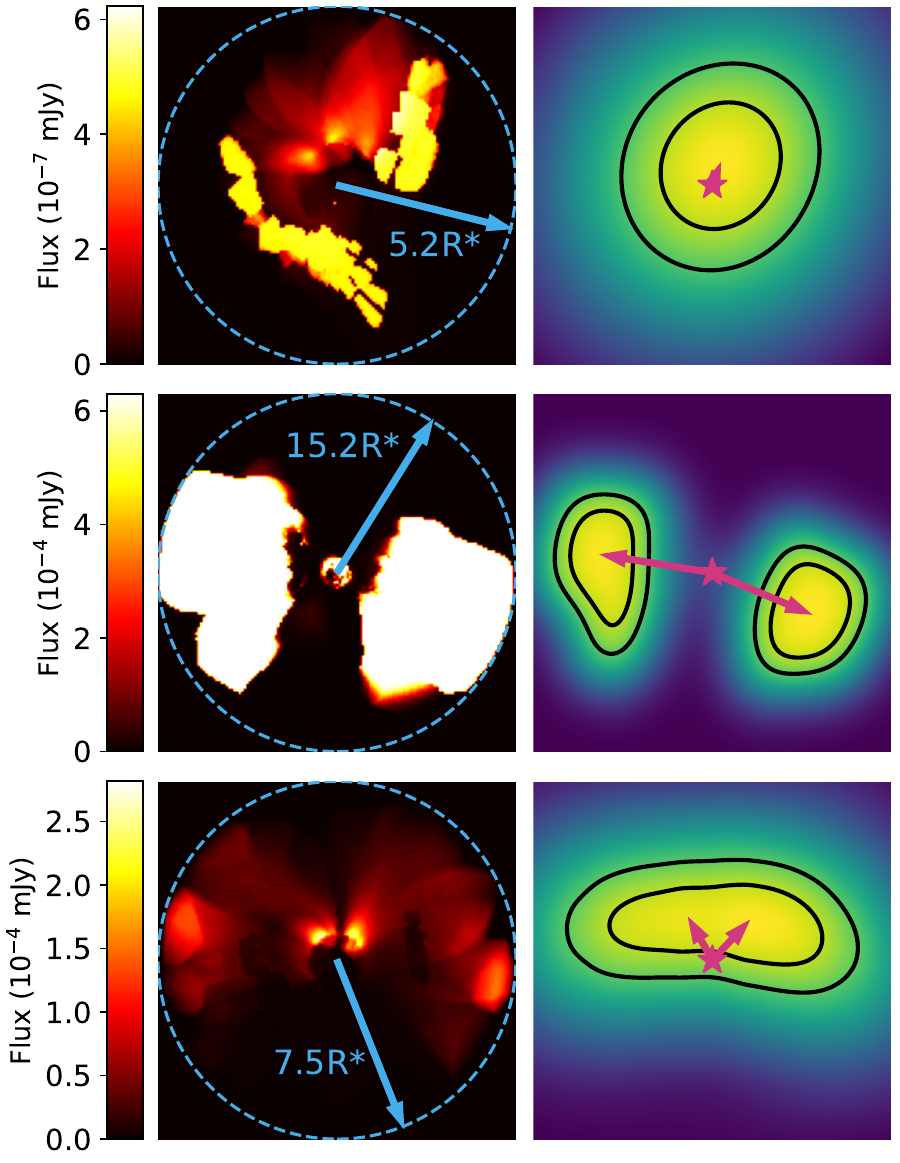}
    \caption{Three examples of synthetic images and their convolved counterparts from different parts of the model parameter space. The synthetic images (\textit{right column}) are all labelled with the model source surface location and individual colormaps. The $75\%$ and $95\%$ contours are overplotted on the convolved images (\textit{right column}); also overplotted are arrows from the the centre of the star to emission peak locations. The model parameters are as follows, \textit{top}: $T_{cor} = 9\times 10^6 \mathrm{K}$, $R_{ss} = 4.2 R_*$, $P_0 = 4\times 10^{-3} \mathrm{Pa}$ \textit{middle}: $T_{cor} = 10^6 \mathrm{K}$, $R_{ss} = 14.2 R_*$, $P_0 = 0.08 \mathrm{Pa}$ \textit{bottom}: $T_{cor} = 5\times 10^6 \mathrm{K}$, $R_{ss} = 6.5 R_*$, $P_0 = 0.04 \mathrm{Pa}$.  \label{fig:poor_fits}}
\end{figure}

Figure \ref{fig:poor_fits} shows example images (high resolution and convolved) from three different parts of the parameter space. The top model has a high temperature corona ($9\times 10^6 \mathrm{K}$), a small source surface ($4.2 R_*$), and low surface pressure ($4\times10^{-3} \mathrm{Pa}$). While in the high resolution image we can see a clear two lobe structure, because of the small scale as compared to the telescope beam size, the convolved image only shows one peak. Additionally because of the low  pressure, the coronal density and thus flux is very low, far below the values observed by \citet{Climent_2020A&A...641A..90C}, and likely so low that it would not be observable if this were the true flux. The  middle model has a low temperature corona ($10^6 \mathrm{K}$), a large source surface ($14.2 R_*$), and mid-range surface pressure ($0.08 \mathrm{Pa}$). While in the high resolution image there is bright emission from just near the star itself in addition to a very bright and obvious lobe to each side, the convolved image shows only the two side lobes. For the bottom model all parameters are in the middle of our range ($T_{cor} = 5\times 10^6 \mathrm{K}$, $R_{ss} = 6.5 R_*$, $P_0 = 0.04 \mathrm{Pa}$). In the high resolution image we can see that while there is emission further out, the brightest emission spots are near the surface of the star, this is reflected in the close location of the two detected peaks in the convolved image, marked pink arrows.

\begin{figure}
 \centering
    \includegraphics[width=\columnwidth]{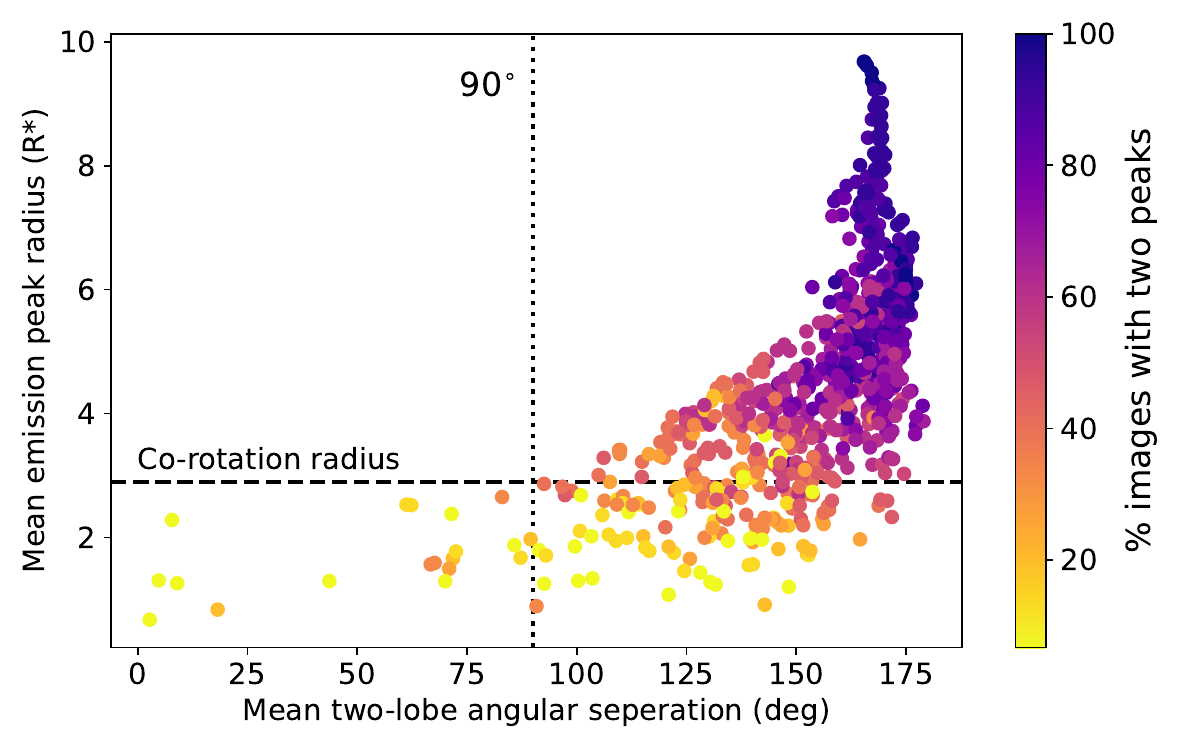}
    \caption{Illustration of the emission lobe characteristics across our parameter space. Mean angular-separation (for all two-lobe images over the rotation period) is plotted against mean lobe radius with the points coloured by percent of the stellar rotation showing two peaks. The Keplarian co-rotation radius, and $90^\circ$ are also indicated. The distribution is such that models most clearly exhibiting the two-lobe structure also show a large angular separation (emission lobes are on either side of the star), and for models with low angular separation the emission originates close to the stellar surface. \label{fig:ang_sep}}
\end{figure}

Figure \ref{fig:ang_sep} quantifies the different geometries we see in the convolved images produced from our models. For each set of model parameters we plot the mean angular separation (as calculated in Figure \ref{fig:t_vs_rss}) and mean emission lobe radius, and colour the points by the percentage of images (across a rotation period) showing two visible emission peaks. We have also marked AB Dor's co-rotation radius ($r_K = 2.6 R_*$), and $90^\circ$ of angular separation. We can see from this plot that all models with mean angular separation $< 90^\circ$ show emission peaks below the co-rotation radius, and a low percentage of the images show two peaks. This tells us that low angular separation occurs when the bulk of the flux originates near the surface of the star, and that for most of the orbit this creates only a single peak in the convolved image (e.g. Figure \ref{fig:poor_fits} \textit{bottom}). With increasing radius for the emission peaks, we also see a corresponding trend to higher angular separations, and higher percentage of two-lobe images. This tells us that when the emission originates higher above the stellar surface, that emission is to either side of the star, and the two-lobe structure is visible through more of the stellar rotation.

\begin{figure}
 \centering
    \includegraphics[width=\columnwidth]{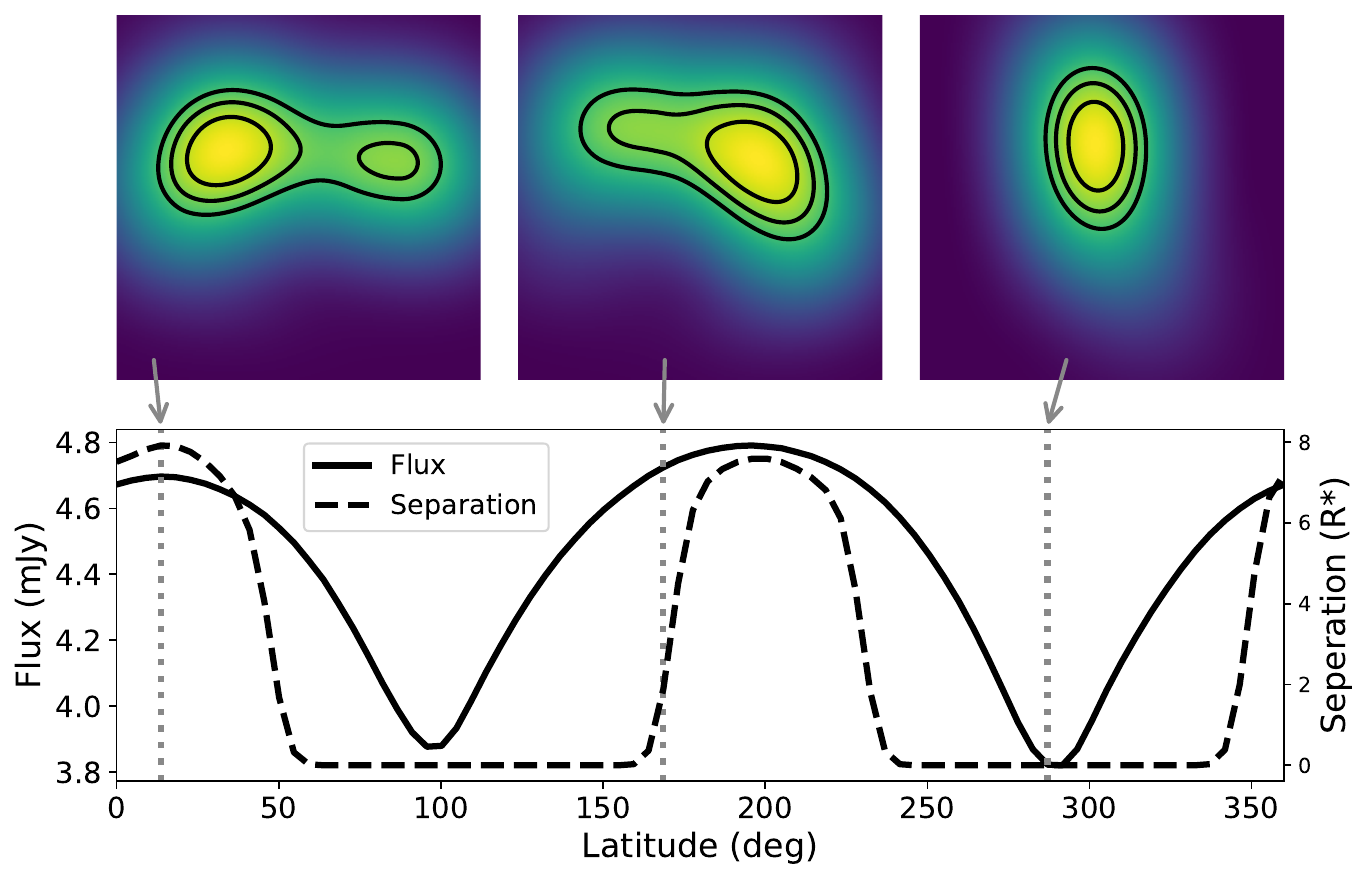}
    \caption{The total flux and peak separation over a stellar rotation period, for our model with parameters shown in Table \ref{tbl:model_parms}. The three images above show the convolved synthetic image corresponding at three example phases: maximum peak separation (\textit{left}), minimum total flux (\textit{right}), and an intermediate phase for both flux and peak separation (\textit{middle}).  \label{fig:phase_rot}}
\end{figure}

While Figure \ref{fig:t_vs_rss}  shows the bulk properties of each model, Figure \ref{fig:phase_rot} shows how one model varies through a full stellar rotation. We can see that the convolved synthetic image moves between a one and two lobe structure, and that the phases of maximum total flux correspond to the phases with the largest peak separation. The total flux varies by $\sim 1 \mathrm{mJy}$, and is clearly related to the geometry of the corona, such that maximum occurs when two emission lobes are clearly visible, and minimum when one is obscured by the other.

\section{Discussion} 

\subsection{Four possible scenarios}

Before discussing our results in detail, we will quickly present the four scenarios presented in \citet{Climent_2020A&A...641A..90C}, and discuss out model's ability to comment on each.

\textit{Close companion hypothesis:} 
As mentioned in the introduction, AB Dor is part of a quadruple star system, that is two gravitationally linked stellar binaries. Thus, \cite{Climent_2020A&A...641A..90C} consider that the two components observed in their data may be interpretable as AB Dor and its low-mass close companion AB Dor C. However, analysing the measured separation of the two visible components over time, assuming they are AB Dor and AB Dor C  yields radial velocity values for AB Dor that significantly higher than prior observations \citep{Climent_2020A&A...641A..90C}. We therefore do not consider this scenario either, modelling AB Dor alone.

\textit{Polar cap hypothesis:} 
In this scenario, the emission seen by \citet{Climent_2020A&A...641A..90C} originates above the magnetic polar region of AB Dor. This is incompatible with our model for all input parameters. Our model produces emission from the magnetic equatorial regions; the polar regions are dominated by stellar wind which contributes negligibly to the overall emission (see Figure \ref{fig:field_lines}). The mechanism proposed to explain this scenario in \citet{Climent_2020A&A...641A..90C} in involves electrons accelerated by flaring activity flowing along magnetic field lines and being emitted as synchrotron radiation over the poles. As discussed in our methods section we modelled neither flaring activity nor synchrotron emission. Thus we can not definitively rule this scenario out, however we find it incompatible with our model, and even if this form of emission were present, one still needs to consider the bremsstrahlung radiation that we do model.

\textit{The helmet streamer hypothesis:} 
As discussed in the introduction, helmet streamers form as particles travel between ``mirror points,'' one point close to the surface of the star, and one directly above it, much higher in the stellar atmosphere. In this scenario, the visible emission lobes are the upper and lower mirror points, meaning that, geometrically, both lobes are to one side of the star, with one considerably further from the star. Figure \ref{fig:ang_sep} shows the distribution of angular separations in our parameter space, and in particular shows that all of our models with the low angular separation necessary to this scenario have emission lobes relatively close to the stellar surface (below the co-rotation radius) and show the two-lobe structure for less than half of their orbit. Thus it is impossible, with our model, to see the $8-10 R_*$ lobe separation that \citet{Climent_2020A&A...641A..90C} observe and also have both lobes on one side of the star. We can therefore rule out this scenario.

\begin{figure}
 \centering
   \includegraphics[trim={0cm 4cm 0cm 6cm},clip,width=.45\columnwidth]{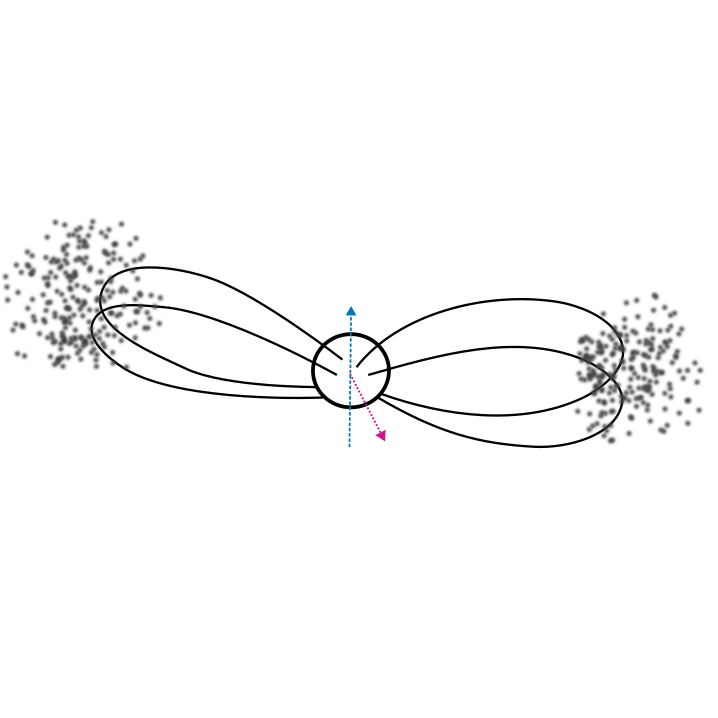}
    \includegraphics[trim={0cm 4.5cm 0cm 1.5cm},clip,width=.45\columnwidth]{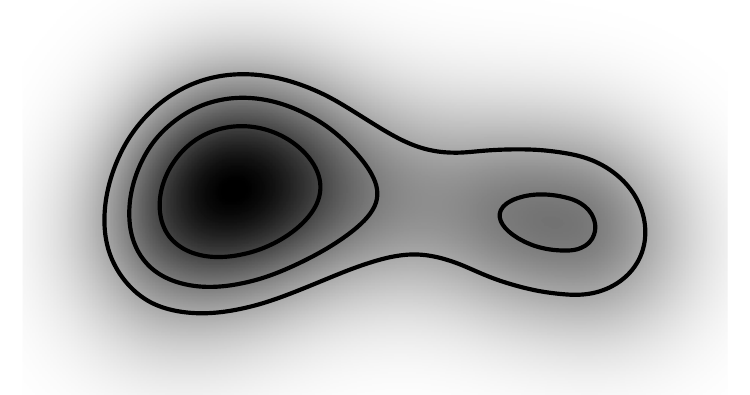}
    \caption{Sketch of the "flaring loops model" from \citet{Climent_2020A&A...641A..90C} next to an example convolved image from a model with the same orientation. The blue arrow marks the rotation axis, and the pink arrow the magnetic dipole axis.
    \label{fig:flaring_loop_diagram} }
\end{figure}

\textit{The flaring loops hypothesis:}
In this scenario the observed emission lobes are the result of emission from the tops of large, closed, magnetic loops in the corona. Figure \ref{fig:flaring_loop_diagram} shows a sketch of this scenario as presented by \citet{Climent_2020A&A...641A..90C} paired with one of our convolved images. Geometrically, this scenario is a match: from our high resolution images we know that the emission is coming from the closed coronal loops, particularly at their peaks, and the majority of our convolved images show the two lobe structure where the lobes are at similar radius to either side of the star. \citet{Climent_2020A&A...641A..90C} suggest, as an underlying emission mechanism for this scenario, magnetic reconnection and interaction events in the upper corona producing a permanent population of accelerated electrons from these frequent but fundamentally transient events. We find that this mechanism is not necessary, rather that a steady state model can produce the requisite flux through simply centrifugal confinement.

\subsection{Coronal Properties}

\begin{figure}
 \centering
    \includegraphics[width=\columnwidth]{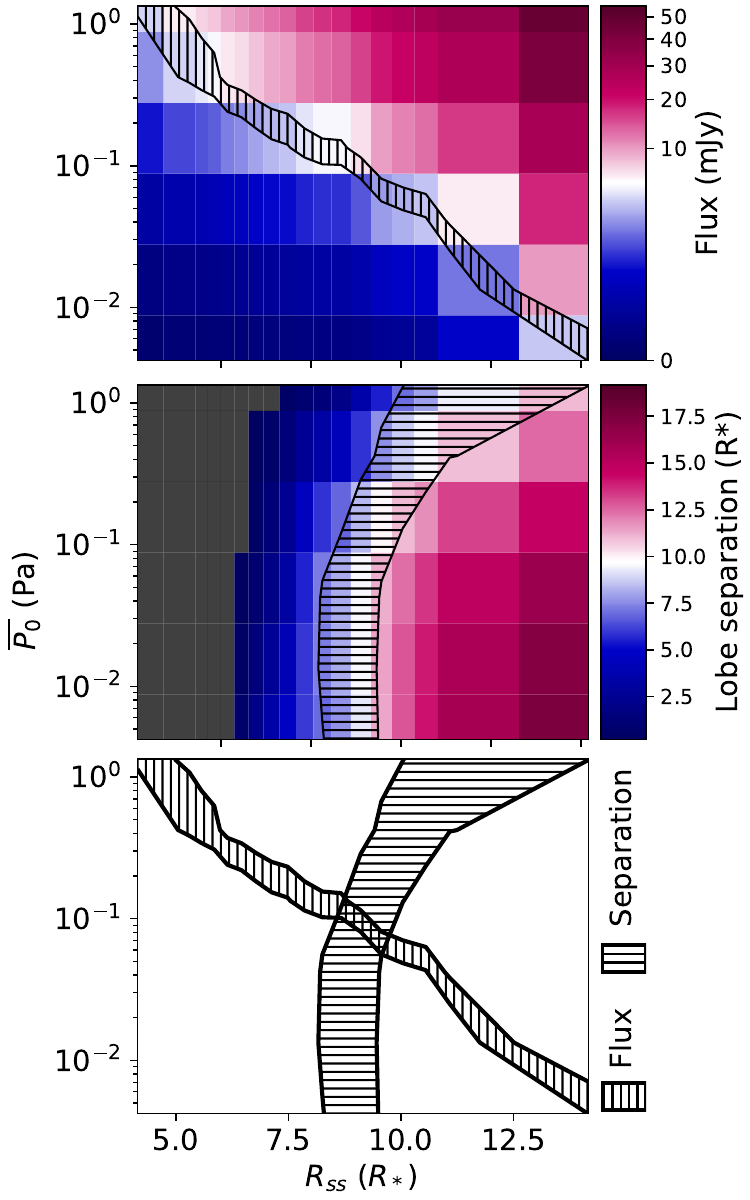}
    \caption{The model mean flux (\textit{top}) and lobe separation (\textit{middle}) as a function of source surface and base pressure for a 9 MK corona. The contours guide the eye to the parameter space that produces the best agreement with the \citet{Climent_2020A&A...641A..90C} results. The \textit{bottom} plot shows both agreement contours, highlighting the small area of overlap.
    \label{fig:fixed_temp_cont} }
\end{figure}

Having established that our model is compatible with the flaring loops hypothesis in \citet{Climent_2020A&A...641A..90C}, we now explore what this can tell us about the characteristics of AB Dor's corona. Figure \ref{fig:fixed_temp_cont} fixes the corona temperature at $9\times 10^6 \mathrm{K}$ and plots source surface vs. base pressure. The \textit{top} and \textit{middle} plots are coloured by total flux and two-lobe separation respectively, with the hatched areas guiding the eye to the values more closely aligned with the \citet{Climent_2020A&A...641A..90C} observations (as in Figure \ref{fig:t_vs_rss}). The \textit{bottom} plot of Figure \ref{fig:fixed_temp_cont} shows just the agreement contours from the two upper plots, shown together so that we can see that they trend in opposite directions and have a tight area of overlap indicating the ``best'' model at this temperature.

\begin{figure}
 \centering
    \includegraphics[width=\columnwidth]{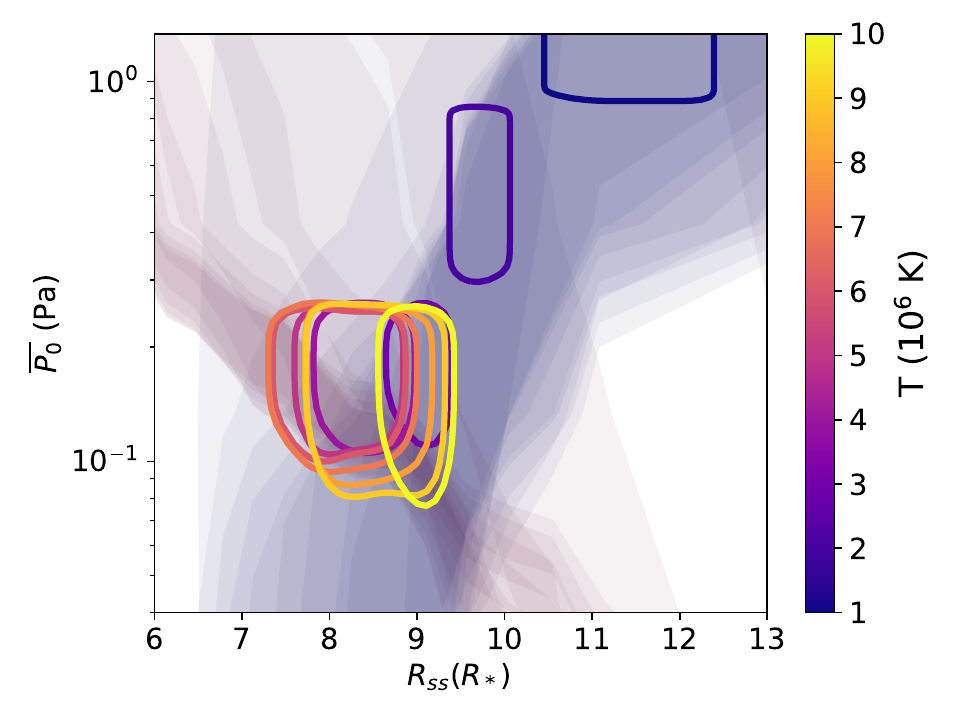}
    \caption{This figure illustrates  the weak dependence on temperature for our model.  The pale blue and pink shading shows the agreement contours (see figure \ref{fig:fixed_temp_cont}) for  mean lobe separation and flux respectively. The coloured contours show the agreement areas for both flux and lobe separation by corona temperature. While the specific values depend on the accuracy of the observations and model, the shape of the contours and overlap are consistent when different target ``accuracy'' values are chosen.
    \label{fig:all_contours} }
\end{figure}

Figure \ref{fig:all_contours} combines the agreement contours shown in the bottom plot of Figure \ref{fig:fixed_temp_cont} for all input temperatures ($10^6 - 10^7 \mathrm{~K}$). The individual flux (pink) and separation (blue) contours are picked out faintly, and the overlapping contours are colour-coded by temperature. We note that there is no contour for the lowest coronal temperature, $10^6 \mathrm{K}$, as there is no overlap in the agreement contours for such a cool corona. We can see that for corona temperatures of $4\times 10^6 \mathrm{~K}$ and hotter the agreement parameter space is very similar; this indicated that for medium to hot coronas, our model is not very temperature sensitive. Prior studies \citep{Hussain_2005ApJ...621..999H, Close_2007ApJ...665..736C} indicate that AB Dor has a hot corona on the order of  $8-10~\mathrm{MK}$, so it is most likely that the AB Dor corona falls into the regime where our model is only weakly dependent on temperature. The base densities for the  region of our parameter space encompassed by the $T = 10^6 - 10^7 \mathrm{~K}$ contours is in the range $n_e = 10^{15} - 10^{16} \mathrm{~m^{-3}}$, which are consistent with values derived from X-ray observations of active stars such as AB Dor (eg. \citet{Ness_2004A&A...427..667N}). This means that without adding further constraints we can assert that AB Dor's source surface is $\sim 8-9 R_*$.

We note that the specific agreement contours shown on these plots are based on the numbers reported by \citet{Climent_2020A&A...641A..90C}, ($4.6\pm0.4 \mathrm{~mJy}$ total flux and $8-10 R_*$ lobe separation), but there is no way to accurately estimate the uncertainty in the comparison because of the complexities and assumptions built into the modelling process. However, examination of the colour gradients in the lobe separation and flux plots in Figure \ref{fig:fixed_temp_cont} reveals that changing the range of values considered to agree with observations moves the location of the agreement contours and therefore the location of overlap, but does not change the shape of the contours or the fact of the overlap. Specifically, increasing the target separation also increases the matching source surface radius (and likewise decreasing), and increasing the target flux also increases the matching base pressure (and likewise decreasing).

\subsection{Exoplanet Environments}
 
\begin{figure}
 \centering
   \includegraphics[width=\columnwidth]{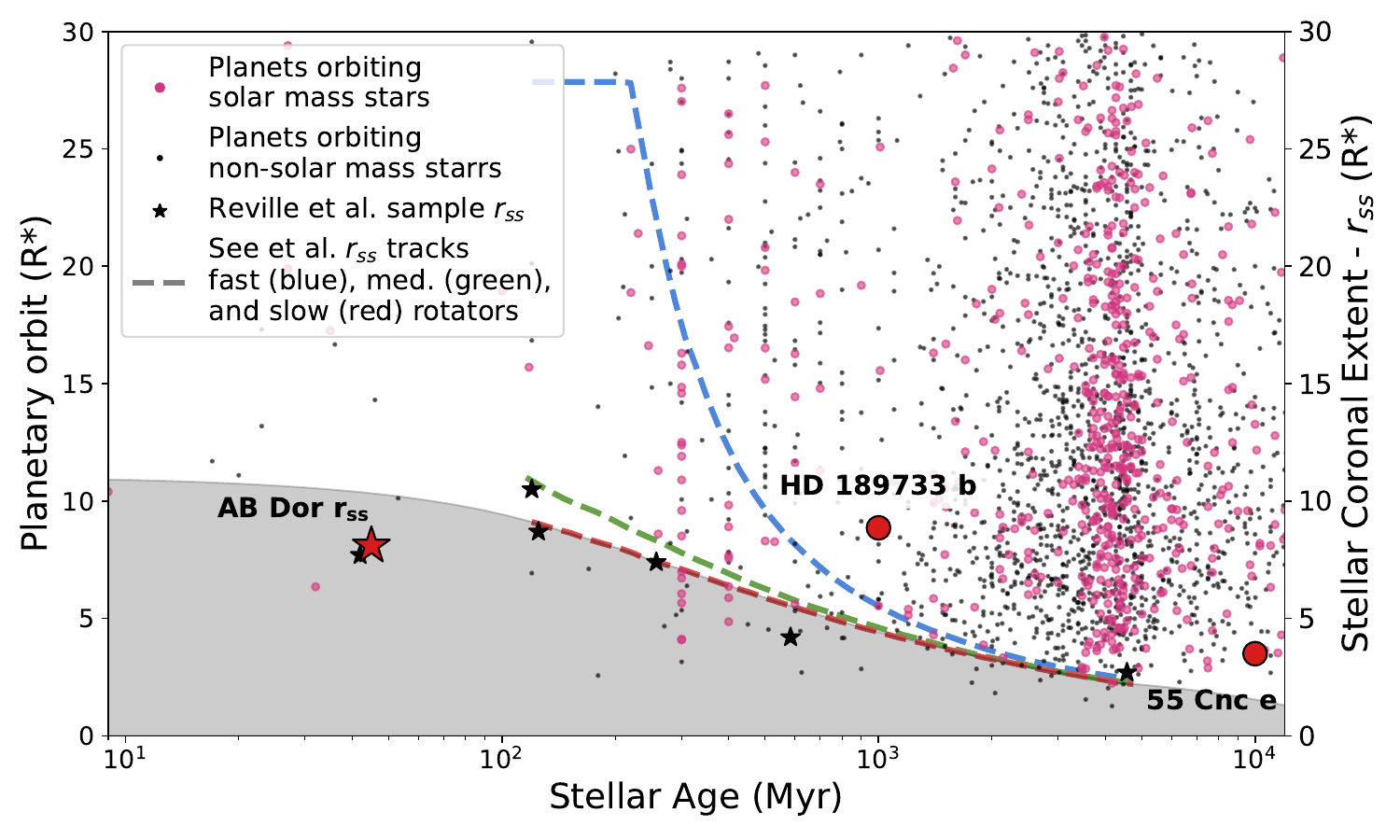}
    \caption{Comparing planetary orbital radii to source surface locations. The black and pink points mark planets orbiting solar mass (pink) and non-solar mass (black) stars, all taken from the NExSI Planetary Systems Composite Parameters Table. We consider stars $0.9-1.1~R_\odot$ to be solar mass. The dotted lines are \citet{See_2018MNRAS.474..536S}'s source surface radii ($r_{ss}$) against age for fast (blue), medium (green), and slow (red) rotators. The shaded region, extrapolated from the slow rotator curve, represents a conservative estimate of the area where planets (at least around solar mass stars) are likely orbiting within the corona. The source surface radii calculated by \citet{Reville_2016ApJ...832..145R} are marked with black stars. Our calculation of AB Dor's source surface, and the orbits of the close-in planets HD 189733 b and 55 Cnc e are labelled.
    \label{fig:planet_orbits} }
\end{figure}

An important implication of the finding that young stars like AB Dor may have far more extended coronae than previously thought is that close-in exoplanets may be orbiting within their star's corona, and if they are not now, they might have in the past. Figure \ref{fig:planet_orbits} illustrates this; on it we plot stellar age versus planetary orbit for all planets in the NASA Exoplanet Archive with known stellar age and planet orbital radius.  Also plotted are theoretical source surface tracks for fast (blue), medium (green), and slow (red) rotators from \citet{See_2018MNRAS.474..536S}, with the slow rotator curve extrapolated beyond the age range discussed in that paper (grey shading). We have also marked the source surface radii calculated by \citet{Reville_2016ApJ...832..145R} for a handful of Sun-like stars with ZDI maps, and note that they align well with the \citet{See_2018MNRAS.474..536S} curves. Even considering the most conservative (slow rotator) evolution track we can see that a handful of planets have a good chance of orbiting within their star's corona.  Furthermore there are a large number of planets in the lower right corner of the plot orbiting older star at radii such that earlier in their evolution they would have been orbiting within the corona of their star, assuming little dynamical evolution as the system evolves. Additionally, the two planets discussed in the introduction as likely orbiting within their star's corona both appear above the source surface curve in figure \ref{fig:planet_orbits}, suggesting that if anything, our source surface estimates underestimate the full extent of closed corona.

This has implications for the interaction between the stellar and planetary atmospheres for close in planets. As discussed earlier in the paper, while the source surface location is a measure of the extent of closed coronal loops, within that imaginary sphere there are both areas of closed corona and stellar wind. Thus, a planet orbiting within this radius spends part of its orbit within the closed corona and part in the stellar wind. Within the closed corona, the thermal pressure is higher due to the hotter and denser environment compared to the stellar wind. The ram pressure however would likely be lower as the gas in the corona does not have the velocity of the wind. What exactly this means for the ability of the planet to retain its atmosphere, and the effect of this changing environment on the planetary magnetic field and surface conditions bears further exploration. Additionally, in this scenario, as the planet orbits the star it would pass across magnetic loops; each time it crosses  a neutral line the location on the stellar surface connected to its movement would change, quite suddenly, to a new location. This behaviour would complicate any search for enhanced activity on the surface of the star related to an orbiting planet, because that signature would jump discontinuously across the surface of the star. And of course, this is even before you account for natural stellar variability as active regions evolve over a period of several weeks.

\section{Conclusions} \label{sec:conclusion}

In this study we have combined two different types of observations of one young star to determine the extent and density of its corona. Both of these parameters are important in studies of magnetic activity and rotational evolution, but they are difficult to determine by any one observational method alone. We have used a surface magnetic map determined from spectropolarimetric observations to generate a 3D model of the magnetic and plasma structure of the young rapidly-rotating star AB Dor. From this model we have determined the free-free radio emission and generated synthetic radio images at all rotation phases. These images typically show a two-lobed structure. By varying the two free parameters of our model (the base coronal pressure and the extent of the closed-field corona) and comparing to the observationally-derived values of the flux density and lobe separation, we were able to:

\begin{itemize}
    \item Reproduce the radio flux (3.7-5.7 mJy) and lobe separation ($8-10~R_*$) of the \citet{Climent_2020A&A...641A..90C} radio imagery
    \item Reject both very small and very large source surface locations, finding the optimal value for AB Dor is $\sim 9 R_*$
    \item Find evidence that \citet{Climent_2020A&A...641A..90C}'s flaring loop hypothesis is the most likely scenario
    \item Reject \citet{Climent_2020A&A...641A..90C}'s helmet streamer hypothesis
    \item Find evidence that the centrifugal confinement of gas at the tops of large closed magnetic field lines  produce the two-lobed radio image structure
\end{itemize}

This study demonstrates the potential power of such a multi-wavelength approach. Coupling both magnetic field mapping (through techniques such as Zeeman-Doppler imaging) with radio images can allow us to measure coronal extents and densities in young solar-like stars.

\section*{Acknowledgements}

We thank the referee for a very thorough and constructive report that has helped us to substantially improve the paper. 

This research has made use of the NASA Exoplanet Archive, which is operated by the California Institute of Technology, under contract with the National Aeronautics and Space Administration under the Exoplanet Exploration Program.

MMJ  acknowledges support from STFC consolidated grant number ST/R000824/1. CEB acknowledges funding from the St Andrews GH Chaplin scholarship. The authors have applied a creative commons attribution (CC BY) licence to any author accepted manuscript version arising.

\section*{Data Availability}

Please contact the authors to request data or software access.



\bibliographystyle{mnras}
\bibliography{Prominences,Solar,Software}




\bsp	
\label{lastpage}
\end{document}